\documentclass{elsart}
\usepackage{graphicx}
\usepackage{graphics}
\usepackage{amssymb}
\usepackage{amsmath}
\usepackage{bm}
\usepackage{lineno}

\newcommand{\be}{\begin{equation}}
\newcommand{\ee}{\end{equation}}
\newcommand{\bea}{\begin{eqnarray}}
\newcommand{\eea}{\end{eqnarray}}
\newcommand{\beal}{\begin{align}}
\newcommand{\eal}{\end{align}}
\newcommand{\bespl}{\begin{split}}
\newcommand{\espl}{\end{split}}

\newcommand{\nslash}{\kern 0.2 em n\kern -0.50em /}
\newcommand{\kslash}{\kern 0.2 em k\kern -0.45em /}
\newcommand{\pslash}{\kern 0.2 em p\kern -0.50em /}
\newcommand{\Sslash}{\kern 0.2 em S\kern -0.50em /}
\newcommand{\Pslash}{\kern 0.2 em P\kern -0.50em /}
\newcommand{\Rslash}{\kern 0.2 em R\kern -0.50em /}

\begin{document}

\begin{frontmatter}

\title{
A scheme for fast exploratory simulation of azimuthal asymmetries 
in Drell-Yan experiments at intermediate energies. The DY\_AB 
Monte Carlo event generator. 
} 

\author{A.~Bianconi}
\address{Dipartimento di Chimica e Fisica per l'Ingegneria e per i 
Materiali, Universit\`a di Brescia, I-25123 Brescia, Italy, and\\
Istituto Nazionale di Fisica Nucleare, Sezione di Pavia, I-27100 Pavia, 
Italy}
\ead{andrea.bianconi@bs.infn.it}

\begin{abstract}
In this note I report and discuss the physical scheme 
and the main approximations used 
by the event generator code DY\_AB. This Monte Carlo code is 
aimed at preliminary  
simulation, during the stage of apparatus planning,  
of Drell-Yan events characterized by azimuthal 
asymmetries, in experiments with moderate center of mass energy 
$\sqrt{s}$ $<<$ 100 GeV. 
\end{abstract}


\end{frontmatter}

\maketitle


\section{Introduction}

Here I discuss the general (physical) scheme of 
the series of event generators DY\_AB, concentrating 
specifically 
on the last version of this code DY\_AB5. 

The event generator DY\_AB5 is a generator of 
dilepton Drell-Yan events\cite{DrellYan70} in hadron-hadron,  
hadron-nucleus and hadron-(partly polarized molecular target) 
collisions. It is aimed at fast preliminary 
simulation of that subset of Drell-Yan experiments, where 

(i) the 
center of mass energy is ``intermediate''  
(from a few to some tenths GeV); 

(ii) the projectile may be any light hadronic species 
(charged pion, proton, antiproton), possibly polarized; 

(iii) the target is in general a molecular species, with 
partial normal polarization of some of its component nuclei; 

(iv) the final leptons present azimuthal 
asymmetries and these asymmetries are the goal of the 
measurement. 

Several experimental proposals have been presented or 
are in preparation in this 
field\cite{panda,assia,pax,compassDY,rhic2}.

The main difficulty of such experiments is the need to 
select regions of the overall phase space where the 
event rates are small (in particular: transverse momentum 
over 2 GeV/c), and where two event numbers (e.g.: before/after 
reversing spin) 
must be compared to identify small asymmetries. It is 
essential to understand from the very beginning which 
overall event numbers are needed to reach a satisfactory 
population of the interesting subregions. 

The here discussed code is aimed at such ``preliminary'' 
investigations, for experimental planning only. Since it 
is based on strong phenomenological components, it is 
not suitable as it is for theoretical analysis at 
quark-parton level. 

Up to date, five (private) versions  
of this code, named DY\_AB1, 2, 3, 4,  
and 5, have been used and tested by the author 
and by other users 
both for phenomenological 
publications\cite{BRMCa,BRMCb,BRMCc,BRMCd,BR_JPG2,AB_06} 
and for exploratory simulations 
aimed at experimental proposals\cite{panda,assia,compassDY} 
(see e.g. \cite{panda_note}, \cite{maggiora1}). 

The latest release DY\_AB5 is 
public\footnote{It may be obtained from the author: 
bianconi@bs.infn.it}.

This code is not a multi-purpose code. 
Its main 
advantages are in its specificity, and are: 
(i) easy insertion of new 
parametrizations for distribution functions associated with 
azimuthal asymmetries, (ii) easy control and modification 
of the code, 
(iii) possibility of simultaneous treatment of events in 
the Collins-Soper reference frame and in the fixed target or 
collider frame, (iv) fast generation of events, 
(v) satisfactory phenomenological reproduction of 
transverse momentum distributions. 

The present note is not the ``readme'' handbook of the code. 
The code itself is normally accompanied by a readme file 
supplying user help. 
Here I discuss the physical scheme used for the 
event generation. This is inspired by the parton model, but 
some simplifications or phenomenological 
parameterizations have been introduced into the standard 
relations for the cross section. 

The reasons behind these simplifications are two: 

(i) This 
code is not aimed at improving the theoretical understanding 
of quark-quark interactions; it is used for reproducing as 
realistic as possible event distributions and associated 
errors, in measures where some gross features of the data  
are already well known, while other ones are largely unknown. 

(ii) At the present stage, the real point is to understand 
whether or not certain measurements will be possible, or 
at which extent they will be possible. This requires a huge 
amount of exploratory simulations, to be run in the smallest 
possible time and with the maximum possible flexibility. 

I hope this presentation clarifies in which frameworks the code 
can be used, or should not be used. 

\subsection{Development notes}

This code has been written in c++ since the first 
version. It began as a toy model Drell-Yan generator, 
aimed at fast exploratory simulation for the 
Drell-Yan measurement within the PANDA 
experiment\cite{panda_note}. After the very first applications, 
the number of options has increased exponentially. 

It was initially used by people of some experimental 
collaborations, in a form that permitted them to 
handle input in a simple way, assuming that they would 
not need touching the code. 
This has shown to be unrealistic. On the other side, 
unrealistic as well 
has been the hope that people could be made able to 
modify the code themselves, without interacting at all 
with the author. And also attempts to organize a 
``once and for all'' form of the code have failed, 
just for the fact that the field is quickly evolving. 
For example, it is difficult to find a ``universal'' 
form for the new distribution functions that one could 
like to insert in the next five years. 

So, the general idea is that apart for a central core 
of classes/functions there is nothing sacred in the 
code structure, and that users 
should be able to modify the code via an as small 
as possible interaction with the author. 

The first versions like DY\_AB1 fully exploited the 
possibility of writing complex hierarchies, offered 
by c++.  In DY\_AB4 this structured form was 
abandoned, but for efficiency purposes massive use was 
made of pointers. DY\_AB4 is well tested, and is 
the most efficient code of this series. A short 
presentation of this code may be found 
in \cite{panda_note}. Its main disadvantage was 
hard readability, since to increase efficiency it 
exploited systematically the 
fortran-style technique of organizing big data 
structures, with functions working on these data  
without explicitly getting them as arguments 
(the ``common'' areas 
in fortran; in c++ the same is obtained via pointers 
to data classes). The absence of explicit arguments 
in the function calls makes the code hierarchy difficult 
to see at first reading. 

DY\_AB5 is less efficient (about 30 \% more time-consuming). 
The advantage is that 
it is much easier to read and modify 
(pointers have mostly disappeared). 
It offers more pre-cooked options as far as 
unpolarized and single-polarization 
Drell-Yan are concerned. 

\section{General theoretical notes}

\subsection{General scheme}

This code typical cycle consists of the following steps:

1) Random event generation in the Collins-Soper frame for  
a specified class of projectile and of target hadrons 
(e.g. negative pion and polarized proton). 

2) Transformation of these events first to the hadronic 
center of mass frame (``collider frame''), and next to 
the fixed target frame. 

3) Loop over events taking into account the composition 
of the (molecular) target in terms of different nuclei. 

4) If required, a number of repetitions of a multi-event 
simulation is performed, possibly with spin reversal. 

5) If required, statistical analysis 
of azimuthal asymmetries is performed, 
calculating averages and fluctuations of the results 
obtained in the 
simulations of step (4). 

Here I discuss the general problems and some 
implementation detail 
connected with steps (1) and (2). 

The exact and complete cross sections on which the event 
generation is based 
may be found in \cite{BR_JPG2}. There, two simulation 
schemes are compared. DY\_AB5 is 
based on what is named ``scheme II'' in ref.\cite{BR_JPG2}. 
An alternative code has been prepared by this author, 
based on ``scheme I'', 
and in \cite{BR_JPG2} the differences in the outcome are 
shown. Also, the reasons are discussed why ``scheme I'' 
is more proper and supposed to take over in the long run, 
but not yet well suited given the present 
state of the art on the phenomenology side. 

Formulas reported in \cite{BR_JPG2} and implemented in 
DY\_AB5 are rather long. 
Here, a simplified version of these relations 
is discussed, to clarify the general cross section 
form, 
the main approximations contained in it, and the way the 
code exploits it. 


In the language of \cite{BR_JPG2} 
``Scheme I'' 
is the parton model cross section for Drell-Yan 
events with most of the necessary details: several 
products of distribution 
functions 
associated with unpolarized and polarized partons and hadrons, 
and with all quark and antiquark flavors, are summed . 

``Scheme II'' exploits some more approximations: (i) A cumulative 
event distribution is factorized out of the sum of all terms. 
This distribution is built by a set 
of simplified parton distribution functions $f$, and is 
phenomenological in the transverse momentum dependence. 
(ii) Generalizing an 
approximation method that may be found e.g. in\cite{Boer99} 
all the least known distribution functions 
$h$ (those associated with azimuthal asymmetries) 
are expressed as ratios $h/f$. (iii) Each such 
$h/f$ term is added to the sum, ``valence-weighted'' by 
ratios like 4/9 etc. 

\subsection{The overall cross section}

The relations discussed in the following may be found, e.g.,  
in \cite{Conway89} and \cite{Anassontzis}. 
The problem is that 
in these two works 
several of such relations assume systematically 
different forms. 
Since this difference is present 
in all the relevant literature, we name these two works 
``ref.A'' and ``ref.B'' 
and systematically report the differences. 

I name ``mass'' $M$ the invariant mass of the lepton 
pair (it is also named $Q$ in some references). 
The indexes ``1'' and ``2'' represent the target and beam  
hadrons.  

The Drell-Yan differential 
cross section can be written in an approximate factorized way, 
inspired by the parton model (see \cite{Field}, chapter 5, and 
refs.A and B): 

\begin{equation}
{d\sigma\over {d\tau dX_F dP_t d\Omega}} \ =\ 
{K(\tau) \over s}\cdot 
\bar S(\tau,X_F)\cdot S'(P_t)\cdot A(\theta, \phi, \phi_s).
\end{equation}

or equivalently in the form 

\begin{equation}
{d\sigma\over {dX_1 dX_2 dP_t d\Omega}} \ =\ 
{K(\tau) \over s}\cdot 
S(X_1,X_2)\cdot S'(P_t)\cdot A(\theta, \phi, \phi_s).
\end{equation}

As customary, $s$ 
is the squared invariant mass of the two colliding 
hadrons. 
\begin{equation}
s\ \equiv\ (E_{CM})^2\ 
=\ 2 m_p (m_p + E_{\bar p,LAB}).
\end{equation}
The scaling assumption means that the only 
dependence of eq(1) or (2) on $s$ should be contained in the $1/s$ term. 

$X_F$ and $\tau$ 
are invariant adimensional variables 
associated with the beam-axis momentum projection, and with the 
virtuality,  
of the virtual photon produced in 
$\bar p$ $+$ $p$ $\rightarrow$ $X$ $+$ $\gamma^*$ $\rightarrow$ 
$X$ $+$ $\mu^+\mu^-$. 
The pair $\tau$, $X_F$ can be substituted by 
the equivalent pair $X_1$, $X_2$.  Then $\bar S$ and $S$ are related 
by a Jacobian factor. 
The definition of $\tau$, $X_F$, $X_1$ and $X_2$ 
is given below, where it requires some discussion. 

$\vec P_t$ is the transverse $2-$dimensional component of the 4-momentum 
of the virtual photon, with respect to the beam axis. 
(it is also named $q_t$, or $Q_t$). 

The angular variables 
$\theta$, $\phi$, $\phi_s$ appearing 
in $A(\theta, \phi, \phi_s)$ are measured in a reference frame where 
the virtual photon is at rest. In this frame $\theta$ and $\phi$ are 
the polar and azimuthal angles of the momentum of one of the two muons, 
while $\phi_s$ is the azimuthal angle of the target spin. 
These variables are discussed 
later, in the paragraph on the angular distributions. 

In the right hand sides of equations (1) and (2) 
$\bar S(...)$ and $S(...)$ 
differ because of the direct substitution 
$\tau$ $= $ $\tau(X_1,X_2)$, $X_F$ $=$ $X_F(X_1,X_2)$, 
and because $\bar S(...)$ $=$ 
$J S(...)$, where $J$ is 
the Jacobian of the coordinate transformation between $d\tau dX_F$ 
and $dX_1 dX_2$.

The coefficient $K$ $=$ $K(\tau)$ is normally named ``K-factor'' 
and is predicted to be 1 in the parton model. 
Actually it is neither 1 nor 
constant (it is $\approx$ 2). Traditionally $K$ contains 
all the parton model violations, which are so kept apart from 
the rest of the cross section. 
Summarizing the PQCD corrections 
into a single $\tau-$depending factor is at a certain extent 
justified for 
$\tau$ far from its kinematic boundaries 0 and 1 
(see \cite{Field}, subsections 5.2, 5.3, 5.5, and \cite{SSVY}), 
and for moderate transverse momenta $<<$ M. At these conditions 
the parton-parton $\rightarrow$ $\gamma^*+X$ cross section 
is dominated by those terms where $X$ subtracts no invariant mass 
from the parton-parton $\rightarrow$ $\gamma^*$ transition 
predicted in the plain parton model.

The above cross section factorization is formally exact within the 
parton model, but 
in the above reported form it hides that 
$S'(P_t)$ is (weakly) dependent on $\tau$ and $X_F$, and that 
$A(\theta, \phi, \phi_s)$ depends on $X_1$, $X_2$, $P_t$, $M$. 

In the code these dependencies have been taken into account. 
The previous equations are however written in such a way 
to focus on the 
$assumed$ variable hierarchy  
that allows for a kinematic separation of 
the $S$, $S'$ and $A$ contributions (see e.g. refs.A and B 
for examples of the experimental procedure to extract these 
terms from incomplete phase space): 
  
(i) $S(X_F,\tau)$ does not depend on $P_t$ and on the angular variables, 
so for any assigned $X_F,\tau$ (or equivalently, $X_1, X_2$) 
it can be calculated from the cross section integrated over all 
the 
$P_t, \theta, \phi$ phase space and summed over spin; 

(ii) $S'(P_t)$ $\equiv$ $S'(X_f,\tau,P_t)$ does not depend 
on the angular variables, so it can be determined by $\theta, \phi$ 
integration plus a sum over spin. 

The function $S'(P_t)$ and $A(\theta,\phi,\phi_s)$
are defined  so that 
\begin{equation}
\int S'(P_t)\ =\ 1,\ \ 
\int A(\theta,\phi,\phi_s) d\Omega_{\theta,\phi}\ =\ 1,
\end{equation}
integrating over all the phase space, for any given $\phi_s$. 
So the total $\sigma$ is just the 
integral of $K(\tau) S(X_1,X_2)$. 

\subsection{The longitudinal term $S(X_1,X_2)$: definitions 
and dangerous ambiguities.}

We can describe the meaning of $\tau$ and $X_F$ by the following 
relations: 

\begin{equation}
\tau\ \equiv\ {M^2 \over s} \approx\ {M^2 \over M^2_{max}} 
\end{equation}
\begin{equation} 
X_F\ \approx\ \Bigg({P^\gamma_z \over {{P_\gamma}_{max}}} 
\Bigg)_{CM} 
\end{equation} 

$\sqrt{\tau}$ is the ratio of the virtual photon virtuality 
$M^2$ to its kinematic maximum $s$, reached in an exclusive 
$\bar{p}p\rightarrow l^+l^-$ annihilation into dilepton. 
$X_F$ 
is approximately the ratio of  
the beam axis component of the virtual photon momentum (in the 
hadron collision CM) to its kinematic maximum. 
The precise definition of $X_F$ is not univoque in the literature, 
as discussed below. Whatever the exact definition, 
$X_F$ and $\tau$ are normally defined as $measurable$ scalar functions  
of the projectile and target four-momenta. 
Alternatively, they can be substituted by their 
combinations $X_1$, $X_2$, whose 
approximate meaning is the ratio of the longitudinal component 
of each of colliding quarks to the parent hadron momentum 
in the reaction CM: 
\begin{equation} 
X_i\ \approx\ 
\Bigg({{(P_z)_{quark}} \over {(P_z)_{hadron}}}\Bigg)_{CM}. 
\end{equation} 

For $X_1$ and $X_2$ several definitions can be found, all approximately
equivalent at large $s$ and $M$. These definitions fall into two
groups: (a) ``theoretical'' definitions, given in terms 
of the (unmeasured) light cone momenta of the colliding quarks; 
(b) ``experimental'' definitions (as in ref.A or B), which express $X_1$ 
and $X_2$ as combinations of the (measured) variables $\tau$ and $X_F$. 
In the ``theoretical'' case, $X_i$ is the ratio of 
the large light-cone component of the $i-$quark momentum to the 
corresponding component of the momentum of its parent hadron. 
For a rigorous theoretical definition of $X_1$ and $X_2$ see e.g. 
\cite{BDR}. 

In the high energy limit experimental definitions are supposed to 
reproduce 
the corresponding theoretical ones, so to access approximately to the 
quark momenta. It must be remarked that this is $not$ 
the situation, in the some portions of the kinematic range 
considered here. 

The definition of $\tau$ is the same in refs.A and B, 
and seemingly ``$\tau$'' means the same thing in all the 
literature on the subject: 
\begin{equation}
\tau\ =\ M^2/s\ \ \ \ (refs.\ A\ and\ B)
\end{equation}

On the contrary, 
for $X_F$, $X_1$, and $X_2$  we have non univoque definitions. 
Ref.A uses 

\begin{equation}
X_F\ =\ {{2 P_L} \over \sqrt{s}}\ \ \ \ (ref.\ A)
\end{equation}
\begin{equation}
X_F\ =\ X_1-X_2,\ \ \  \tau\  =\  X_1 X_2 \ \ \ (ref.\ A). 
\end{equation}

Ref.B uses 
\begin{equation}
X_F\ =\ {{2P_L} \over {\sqrt{s}(1-\tau)}}\ \ \ \ (ref.\ B)
\end{equation}
\begin{equation}
X_F\ =\ (X_1-X_2)/(1-\tau), \ \ \tau\ =\ X_1 X_2 \ \ (ref.\ B).
\end{equation} 

The definitions of ref.A are easier to use and more common 
in the literature, 
so the code sticks to them. 
With them it is necessary to take care with the kinematic 
limits: $\vert X_F \vert_{max} $ $<$ 1. 

When comparing 
differential cross sections referred to the variables $X_1, X_2$ 
or $X_F, \tau$, 
jacobian conversion factors are necessary. 

The differential cross sections of equations (1) and (2) enjoy complete 
scaling properties for the $\bar{p}p$ case, 
while 
in the $\pi p$ case a mass-dependent\cite{BergerBrodsky79} term 
is introduced by ref.A, it is important for large $X_\pi$ 
and considered in the code. 

Full scaling means that $(P_t,\theta,\phi)-$integrated cross sections only 
depend on the $X-$set variables, apart for an $s$ dependence confined to 
the $1/s$ term. 
Compared data of ref.A (250 GeV/c beam energy) 
and ref.B (125 GeV/c) on $\pi-p$ DY scattering show approximate scaling. 

$K$ is assumed to be a function of $\tau$ in ref.A (and in the 
code), while most 
experiments (see the DY database from\cite{HEPDATA})  
use it as a constant normally $\approx$ 2. 
For large $\tau$ values, the data by ref.A show that this 
dependence is relevant, and the results of the calculation 
by \cite{SSVY} support this point. The largest part of the 
events concentrate at the lowest part of the involved 
$\tau-$range (wherever it begins), and this may make it difficult 
to scan a large $\tau-$range, to establish precise dependencies 
for $K$ on $\tau$. 

As remarked in ref.B, the choice 
of the $K$ value depends on the choice of the normalization for 
the quark distribution functions, which is not univoque. Ref.B 
reports a detailed and systematic 
discussion of the different normalization methods and of the consequent 
changes in the values of the distribution function parameters 
and of $K$. This can be exploited, with the warning of using 
distribution functions according to the notations of ref.B (see below). 

In the default version of the code DY\_AB5, $S(\tau,X_F)$ 
has been reconstructed using the parameterized form given in 
appendix A of ref.A, together with the kinematic definitions and 
structure functions contained 
in the main text of ref.A. 

This allowed me to fit 
$\pi^--$Tungsten DY differential cross sections reported by that 
experiment at 252 GeV/c. 

When notations of the two works are conformed each other, 
the distribution functions fitted from ref.A allow to reproduce 
reasonably $\pi^--$data reported by ref.B at 125 GeV/c. 
To reproduce the $\bar p-A$ DY data of the same 
experiment, proton quark distribution functions must be used as  
antiproton antiquark distribution functions, and then the 
reproduction 
is reasonable. 

In both references and everywhere in the literature 
$S(X_1,X_2)$ has the form

\begin{equation}
S(X_1,X_2)\ =\ G(X_1,X_2) \Sigma_i \bar F(X_1) F(X_2)
\end{equation}

where $G(X_1,X_2)$ is a kinematic factor proportional to 
$(1/X_1X_2)^2$ (ref.A) 
and $1/X_1X_2$ (ref.B). 
In general 
its exact form depends on notations and changes from paper to 
paper. The exponent ``1'' or ``2'' in the $1/X_1X_2$ factor 
is very important 
because it indicates whether the distribution functions $F(X)$ 
must be read as $F(X)$ or as $X F(X)$ (see below). 

$\bar F$ and $F$ are linear 
combinations of the $\bar q/q$ main distribution functions 
$u(X)$, $d(X)$, $s(X)$. For these we have 

\begin{equation}
X\cdot u(X)_A\ =  u(X)_{B}, \ \ 
X\cdot d(X)_A\ =  d(X)_{B},\ etc.
\end{equation}

So, e.g., the normalization $\int X dX (u+d)$ $=$ 0.34 in ref.A
becomes $\int dX (u+d)$ $=$ 0.34 in ref.B. The $\alpha$ parameter 
appearing in the typical parameterization $u(X)$ $=$ 
$X^\alpha (1-X)^\beta$ changes by one unit passing from ref.A to 
ref.B, and so on. 

The important remark is that this ambiguity $q(X)$ 
$\leftrightarrow$ $Xq(X)$ is present throughout all the literature 
on the subject, not only in these works. This is 
a very delicate point and 
must be taken 
into account whenever new terms are added to the code. 

\subsection{The $P_t$ dependent $S'(P_t)$.} 

The traditional parton model literature is built on the 
collinear approximation. So, for the $P_t-$dependent parts 
one must rely on phenomenological fits. 

Experiments A and B did not impose a low-$P_t$ 
cutoff, with the consequence 
that their small $P_t$ data show a completely different qualitative 
behavior. Measured values of the 
function $S'(P_t)$ can be seen e.g. in ref.A figs.23 and 25 
($\pi+p$ case), or in ref.B fig.9 ($\pi+p$ and $\bar p+p$ cases). 
Since azimuthal asymmetries are very small for $P_t$ 
$<$ 1 GeV/c, this difference is not relevant for the purposes 
of planning experiments on azimuthal asymmetries in Drell-Yan. 
For pions the default option is the distribution used in ref.A. 

The code however offers a series of alternatives. The class that 
handles all $P_t$-distributions is PT2. The use of a specific 
distribution requires subclassing. 

Some options are already present in the code DY\_AB5: 

PT2\_Old : public PT2 is further subclassed into three 
possibilities: 

(1) The distribution by Conway et al\cite{Conway89} to 
reproduce $\pi^--$tungsten at 250 GeV/c. 

(2) The one by J.Webb\cite{Webb} (E866 collaboration) for proton-nucleus 
at 800 GeV/c. 
hep-ex/0301031

(3) The one by Chang\cite{Chang} (E866 collaboration) for the 
same measurement of J.Webb  
at the $J/\psi$ mass. 

(4) The class PT2\_Simple\_Asym : public PT2 of the form 
$NP_t^n/(P_t^2+P_o^2)^m$. This shape 
is useful for the 
azimuthal asymmetry terms (see below). 
 
The most relevant features 
of the measured $S'(P_t)$ 
are (i) a not too strong dependence on $X_1,X_2,s$, (ii) 
for $P_t$ $<$ 2 GeV/c the distribution is not steeply decreasing 
(in case ref.A it is increasing up to 0.5 GeV/c); 
For $P_t$ $>$ 2 GeV/c it decreases steeply (but with power law, 
not exponential); (iii) the average $P_t$ is near 1 GeV/c, 
and as well known (see e.g. \cite{Field})  
it is larger than in lepton-induced DIS 
and in hadron-hadron semi-inclusive meson production. 

In the preliminary simulation of a Drell-Yan experiment 
on azimuthal asymmetries,  
a good phenomenological shape of 
$S'(P_t)$ is a key success factor, because 
measured and/or predicted leading-twist azimuthal asymmetries 
do increase at increasing $P_t$, obliging the experiment to 
select events at a as large as possible $P_t$. 
However, due to the very fast 
decrease of $S'(P_t)$ for $P_t$ $>$ 2 GeV/c, the choice of a 
too large $P_t$ 
cut-off can make a measurement prohibitive 
because of fast falloff of the 
event rates at large $P_t$. 

Ref.A reports an explicit parameterization for $S'(P_t)$ 
(relative to $\pi-$induced Drell-Yan). Ref.B reports  
data and some models for the $P_t$ distributions in both $\bar p p$ 
and in $\pi p$ DY. In the region $P_t$ $>$ 3 GeV/c error bars are 
too big to draw any conclusion, but for $P_t$ up to 3 GeV/c 
$\pi p$ and $\bar p p$ $P_t-$distributions are very similar. 

For pion and antiproton projectiles I did not modify the 
parameterization 
inherited from appendix A (pion) 
of ref.A and from ref.B ($\bar{p}$). 
The pion one is $not$ scale-independent. 
It depends explicitly on $M$ $=$ $s\tau$, and produces a 
slow increase of the average $<P_t^2>$ at increasing 
$s$ and constant $\tau$. 

For proton projectiles, the parameterization by J.Webb\cite{Webb} 
is probably more recommendable. 

\subsection{Isospin/flavor composition.} 

Up to a few years ago the 
models on the functions associated with azimuthal asymmetries  
didn't go to such details like 
the $Z/A$ composition of the target. For this reason, the 
first code DY\_AB1 did not care isospin/flavor matters. 
After Hermes and Compass results, some flavor and isospin-dependent 
parameterizations for single-spin asymmetries have 
appeared (see 
e.g.\cite{BRMCc,Torino05,VogelsangYuan05,CollinsGoeke05}), 
obliging the codes since DY\_AB2 onwards to take these problems 
into account. 

On the experimental 
side the Z/A composition is important for another reason: 
it determines 
the effective dilution factor of the target polarization.  

DY\_AB5 takes into account 
events coming from separate pieces of a molecular 
target, taking the individual dilution factor of each nucleon 
into account. 
So to reproduce a $NH_3$ target with 85 \% polarized $H$, 
one may arrange code parameters 
so to require about 4 events  
on an unpolarized proton or neutron and one event on a 
polarized proton. After this, any specific sorted event, e.g. 
$\pi_--$neutron, will actually need to be translated 
into a $\bar{u}u$,  
a $\bar{d}d$ or a $\bar{s}s$ event. 

In DY\_AB5 this is taken into 
account by $X-$averaged weighting factors. The criterion and 
the weights for the relevant cases are discussed in detail 
in ref.\cite{BR_JPG2}. There, also the errors associated 
with this technique are shown, in 
``approximate vs exact'' scatter plots. 

The point is that a sum of the kind 
\begin{equation}
\sum_{flavor} \Big( S_{leading}[1 + S_{asymmetry}/S_{leading}] \Big)
\end{equation} 
where each term is flavor-specific, 
is approximated in the form 

\noindent 
\begin{equation}
S_{leading} \Big(1 + 
\sum_{flavor} W_{flavor} [S_{asymmetry}/S_{leading}] \Big),
\end{equation}
where the leading term is common, 
the weights are constant (and of course depend on the 
projectile-target hadron pair for the sorted event), the 
terms $[...]$ are flavor-specific functions where it is not 
possible to separate the numerator from the denominator. 

Weight factors 
are needed to compensate the fact that e.g. the asymmetries 
associated with $\bar{d}d$ collisions have scarce relevance 
in a process like $\bar{p}p$ or $\pi_- p$, while they have more 
in $\pi_+n$. 
As observed in \cite{BR_JPG2}, the discrepancies between the 
``exact'' scheme and the constant-weight scheme 
are relevant when both the colliding hadrons 
have small $X$. 

In practice, planned experiments will 
try to stay as far as possible from this region, since 
common belief is that transverse spin effects 
are small there. 
In addition, although it would be better to sort 
events according to the former scheme, phenomenological 
parameterizations do normally extract 
the ratios $[S_{asymmetry}/S_{leading}]$ according to the 
latter scheme. Paradoxically, as remarked 
in \cite{BR_JPG2}, this makes the latter scheme more proper 
than the former for a simulation. 

So, although the author of this note 
has prepared since long a ``DY\_AB6'' code working according 
with a more exact scheme (the one used for the 
simulations of \cite{BR_JPG2}), such a code is unlikely to 
be useful for still some time. 

\subsection{Angular distributions} 

Angles $\theta$, $\phi$ and $\phi_s$ in the function 
$A(\theta, \phi, \phi_s)$ are measured in a reference frame 
with origin in the center of mass of the muon couple. 
In other words, the origin of this frame coincides with the virtual 
photon position. 
The axes of this frame can be oriented in several ways. One 
leads to the so-called
Collins-Soper frame\cite{CollinsSoper77} (CS), 
with $\hat z$ axis parallel to the difference of 
the momenta of the projectile and of the target $nucleon$. 
The transverse axes are oriented so that the $xz$ plane contains the virtual
photon momentum. 
Other common alternatives put the $\hat z$ axis along 
the beam or target direction in the lepton CM. 

The DY\_AB5 code sorts events in the CS frame. 
This fact becomes relevant when events are transformed 
to the fixed target or to the collider frame (see 
related section below). 

In the CS frame, 
$\theta$ and $\phi$ are the angles formed by $one$ of the muons 
($\mu^+$ from now onwards), and $\phi_s$ 
is the target spin orientation. 

For a $qualitative$ understanding of the kinematics for most (not all) 
events, one may imagine a virtual photon with transverse momentum 
not much larger than 1 GeV, and $\vert X_F\vert$ not too close to 
zero, so that the longitudinal momentum of the photon is larger than the 
transverse momentum. 
Then the CS $z$ axis is roughly parallel to the collider one. Also the 
CS $xy$ plane 
is roughly parallel to the collider $xy$ plane, but the 
CS $x$ and $y$ axes are randomly rotated by an angle 
$\phi_{coll}$ with respect to their collider configuration. This 
$\phi_{coll}$ is the angle between the $xy-$component of the 
virtual photon and the $x$ axis in the lab. The $x$ axis of 
the CS frame coincides physically with the transverse component of the 
virtual photon 3-momentum. Then the transverse proton spin, which 
is fixed along the $x$ axis in the collider frame,  lies 
on the CS $xy$ plane with angle $\phi_s$ $=$ $-\phi_{coll}$. 

These approximations are not used in the code, 
but can be useful to understand 
the meaning of the employed angles, and to have an idea of the 
distribution of useful events in a collider frame. 

In the Collins-Soper frame 
the angles $\theta$ and $\phi$ are polar and azimuthal angles of 
the momentum of one of the two leptons and are randomly distributed.  
Also the spin angle $\phi_s$ 
is randomly distributed in the CS frame. In absence of information on 
two of these angles, the third one is homogeneously distributed  
over all of its phase space. However, altogether 
their distributions are correlated by 
$A(\theta, \phi, \phi_s)$ in the cross section. 

In the simulation events, are initially flat-sorted with respect to 
all the kinematic variables $X_1$, $X_2$, $P_t$, $\theta$, $\phi$, 
$\phi_s$. The sorted events are accepted/rejected according to the 
cross section expressed by eq.(1) 
where, at the level of single spin experiments,  

\begin{equation}
A(\theta, \phi, \phi_s) \ =\  
1\ +\ cos^2(\theta)\ +\ 
{{\nu(X_1,X_2,P_t)} \over 2} sin^2(\theta)cos(2\phi) 
\ +\ 
\nonumber
\end{equation}
\begin{equation}
+\ 
\vert \vec S_2 \vert B( X_1, X_2, P_t, M,\theta, \phi, \phi_s ) 
\end{equation}

The unpolarized asymmetry measured in ref.A is contained in 
the $\nu...$ term. 
Single spin asymmetries arise from the $\vert \vec S_2 \vert B(...)$ 
term.  Two origins for such terms have been 
considered here, corresponding to different origins for the azimuthal 
term. 

The Sivers\cite{Sivers} asymmetry considers a term of the form 

\begin{equation}
B = 2 {m_p \over M} sin(2\theta) sin(\phi-\phi_s) H_a(X_1,X_2)
\end{equation}

while the 
Boer-Mulders\cite{Boer99,BoerMulders98} 
asymmetry is of the form 

\begin{equation}
B = - {1 \over 2} \sqrt{\nu \over \nu_{max}} sin^2(\theta) 
sin(\phi+\phi_s) H_b(X_1,X_2)
\end{equation}

Any of the above azimuthal asymmetries (unpolarized, 
Sivers, BMT) can be disentangled from the other two 
by a suitably weighted $\phi$ integration. 
The ``statistical analysis'' option of the code offers this 
possibility.

\subsection{Parameterizations for the nonstandard 
distribution functions}

As above written, the leading distribution functions 
are bypassed by assuming a phenomenological 
(correctly behaving and scaling in a wide range of kinematics) 
form for the ``standard'' event distribution, i.e. for the 
$(\theta,\phi)-$averaged part of the event distribution. 

For the 
``nonstandard'' terms (i.e. those ones that produce 
nontrivial angular distributions in the CS frame) 
the code includes as a first possibility some 
phenomenological parameterizations 
that have become recently available in the literature, 
and as a second option  
the possibility to chose freely the parameters 
for simple pre-determined shapes. 

The functions associated with the 
nonstandard terms have been put in places where it is 
easy to find them (in the file c\_dy\_master.cpp). 
So, if one wants to change radically the shape of 
these functions it is possible to do it. 
The code has been written assuming that 
any potential user could be interested in adding 
supplementary terms to this set. In other words, I
assume that in the next ten years there will be 
no special reasons to change the ``standard'' part 
of the cross section, while updates and news on the 
asymmetry side will be frequent. 

Although it is possible to add new parameterizations, 
respecting some formats makes things easier. Here I 
would like to discuss these formats. 

1) Let us write the $A$ term in eq.(17) in the 
most general form

\begin{equation}
A(......)\ \equiv\ 1\ +\ cos^2(\theta)\ +\ \sum
\Big(
F(X_1,X_2,P_t)
F'(\theta)F''(\phi,\phi_s)
\Big)
\end{equation}

It must be reminded that according with the scheme 
adopted in this code each $F$ term is the $ratio$ between a 
term associated with a given kind of angular asymmetry 
and the ``standard'' term, that carries with itself the 
$1+cos^2(\theta)$ angular dependence. Since most of 
the available parameterizations give directly such a 
ratio, this is the most convenient form, as earlier 
discussed in this work. 

2) The code assumes factorization between 
longitudinal and transverse degrees of freedom, 
and between terms 
coming from projectile and target, in the functions 
terms $\nu$ and $B$ in eq.(17). 
These functions have the form 

\begin{equation}
F(X_1,X_2,P_t)\ \equiv\ f_1(X_1)f_2(X_2)f_t(P_t),
\end{equation}

3) For the longitudinal components $f_i(X_i)$  
the code assumes the form 

\begin{equation}
f_i(X)\ \equiv\ N X^\alpha(1-X)^\beta
\end{equation}

To insert completely different functions is possible 
but it is of course much easier to change three 
parameters for any quark flavor. 

4) The transverse term is 

\begin{equation}
f_t(P_t)\ =\ 
{
{
\int d^2 k_1 d^2 k_2 
[g_1(\vec k_1)g_2(k_2)]_{asymmetry}\delta^2(\vec P_t - \vec k_1 - \vec k_2). 
}
\over
{
\int d^2 k_1 d^2 k_2 
[g_1(\vec k_1)g_2(k_2)]_{standard}\delta^2(\vec P_t - \vec k_1 - \vec k_2). 
}
}
\end{equation}

\noindent
where as previously reminded, the numerator is the really 
asymmetric term, the denominator is the leading standard 
$P_t-$dependence. 

5) For the transverse part it is 
responsibility of the user to introduce directly the 
convolution of the two transverse momentum 
distributions corresponding to the colliding partons. 
In other words, the code assumes that one directly 
introduces $f_t(P_t)$ in eq.(20).

The last constraint is not a true constraint, since all the 
parameterizations known to me employ gaussian 
shapes, whose convolution may be easily computed in 
analytical way. 
On the other side, this means spared execution time for 
the code. 

7) Frequently, the result of eq.(23) 
$f_t(P_t)$ will assume the form 
that I name ``$f_{simple}(P_t)$'': 

\begin{equation}
f_{simple}(P_t)\ \equiv\ 
NP_t^n/(P_t^2+P_o^2)^m. 
\end{equation}

\noindent
Because of the common use of Gaussian distributions, 
this form for the ratio $f(P_t)$ is quite common. 
So the code gives, among other options, this one. 

To implement the calculated $P_t-$convolutions, or to insert 
phenomenological ones (including the ``standard'' term $S(P_t)$ in 
eq.1) the code offers a class PT2 that 
may be subclassed two ways: 

(i) Exploiting the class PT\_Simple\_Asym : public PT2 
to directly insert the parameters $N$, $P_o$, $n$, $m$, 
into the form $NP_t^n/(P_t^2+P_o^2)^m$. 

(ii) Subclassing the class PT2 by another user-assumed 
distribution. As above discussed, the code itself offers 
several examples of this procedure,  
i.e. all the relevant alternatives offered for the 
$P_t$-dependence of the leading ``standard'' term. 
 
The already present $default$ parameterizations correspond, 
for the Boer-Mulders effect, to the 
$x-$independent $\nu(k_t)$ 
function given in \cite{Boer99}, for the Sivers effect to 
the two alternatives \cite{BRMCc,Torino05}. 

For the $k_t-$unintegrated Transversity distribution 
the default $(N,\alpha,\beta)$ parameters 
are $(1,0,0)$ and no predefined set was considered. 


\section{Interesting Plots}

In the following, some collections of simulated Drell-Yan events 
are compared with the measured distributions, for negative 
pions on Tungsten. At the kinematics of interest for DY\_AB5 
the two experiments with far the best statistics 
in the mass range 4-9 GeV/c$^2$ are E615 at Fermilab and NA10 
at CERN. The code DY\_AB has been based on the scheme and 
parameters given by E615 in \cite{Conway89}, 
however it 
works equivalently with NA10 data\cite{NA10_194,NA10_286}, 
coming from nearby kinematic regions.

\subsection{Comparison with experimental data: E615}

To produce fig.1, DY\_AB5 has sorted 100,000 Drell-Yan events 
for negative pions 
with beam energy 252 GeV on a Tungsten target, in 
the dilepton mass range 4-7 GeV/c$^2$. From this set, 
I extract the subset 
of events with $\sqrt{\tau}$ in the range 0.254-0.277 
(mass from about 5.5 to 6 GeV/c$^2$). The distribution of 
these events with respect to $x_F$ may be compared with the 
cross sections given in \cite{Conway89}, table VI.

\begin{figure}[ht]
\centering
\includegraphics[width=9cm]{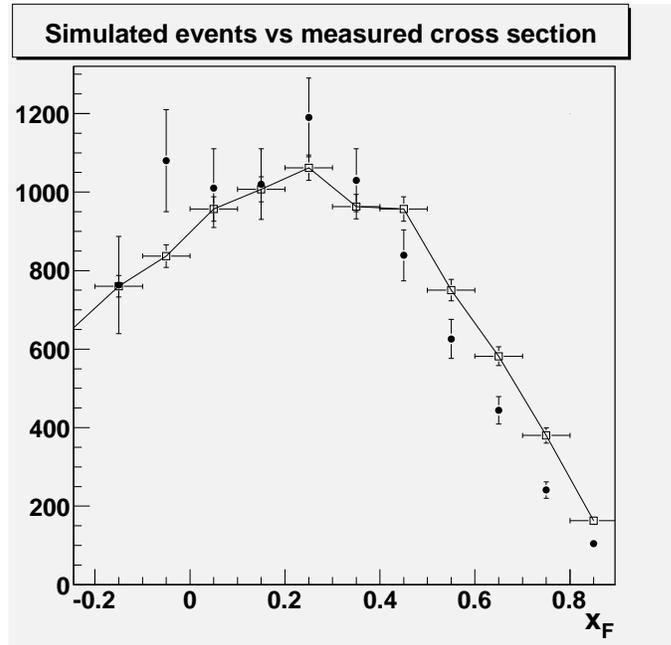}
\caption{
Filled squares: E615 data for 
$\sqrt{\tau}$ in the range 0.254-0.277. 
Empty squares with joining line: Simulation (see text). 
\label{fig:X1}}
\end{figure}

\begin{figure}[ht]
\centering
\includegraphics[width=9cm]{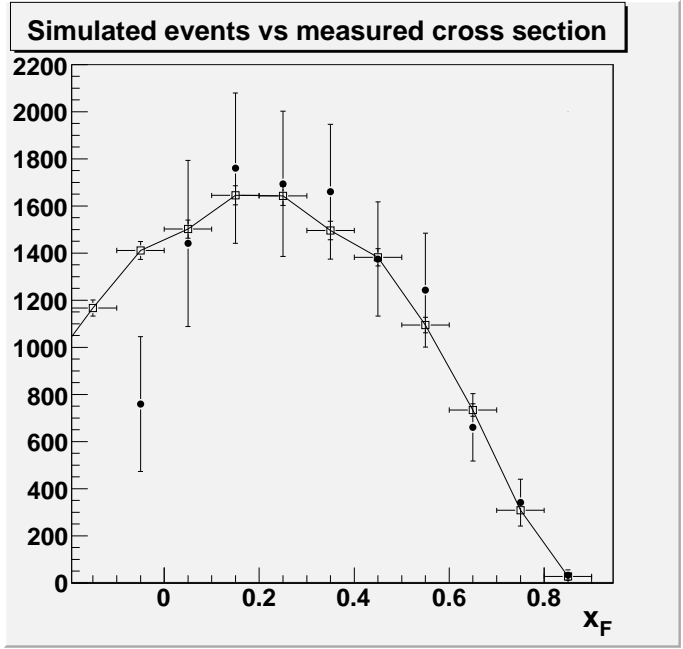}
\caption{
Filled squares: E615 data (252 GeV pion beam) for 
$\sqrt{\tau}$ in the range 0.392-0.415. 
Empty squares with joining line: Simulation (see text). 
\label{fig:X2}}
\end{figure}

\begin{figure}[ht]
\centering
\includegraphics[width=9cm]{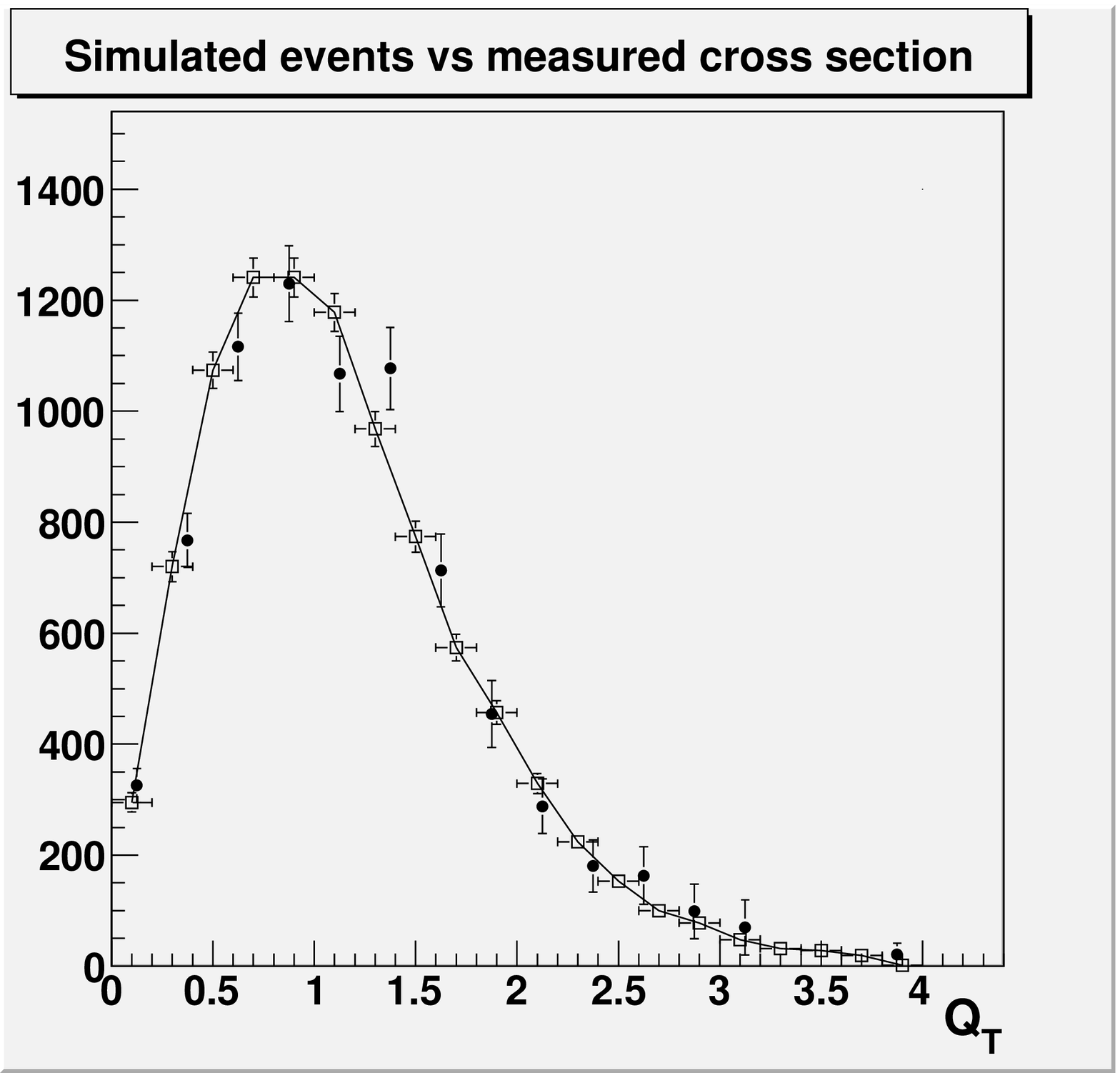}
\caption{
Filled squares: E615 data for 
$x_F$ in the range 0-0.1. 
Empty squares with joining line: Simulation (see text). 
\label{fig:X3}}
\end{figure}

\begin{figure}[ht]
\centering
\includegraphics[width=9cm]{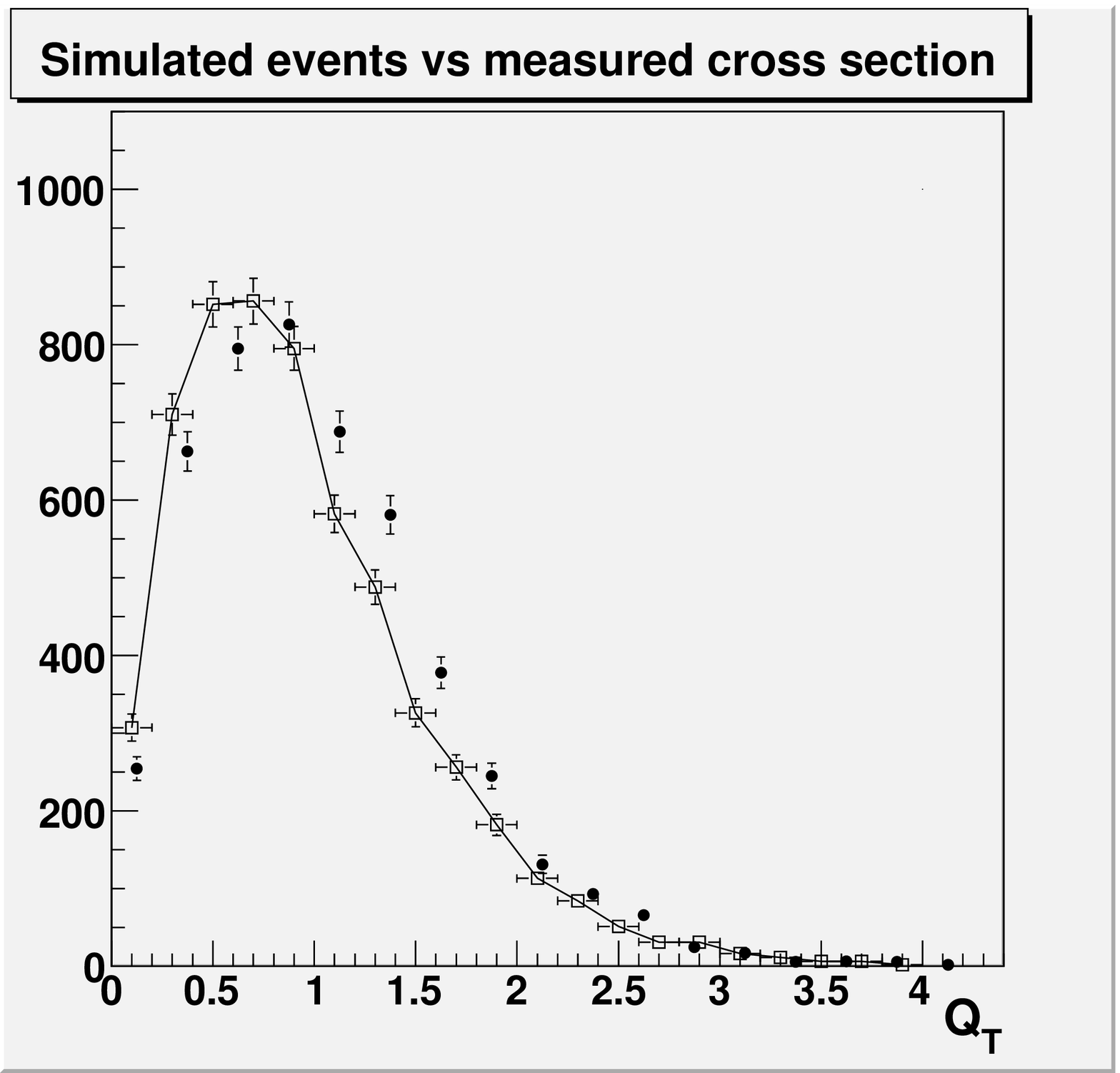}
\caption{
Filled squares: E615 data for 
$x_F$ in the range 0.6-0.7. 
Empty squares with joining line: Simulation (see text). 
\label{fig:X4}}
\end{figure}

In fig.1 the empty squares joint by a line are the event 
numbers sorted by DY\_AB5. Horizontal bars represent the 
size $\Delta x_F$ $=$ 0.1 of each bin, vertical bars the 
statistical error $\sqrt{N}$. The full black squares with 
no horizontal error bar are the numbers from table VI 
of \cite{Conway89}, rescaled by a common constant 
factor so to transform cross section values into 
expected values for the bin populations. 

Sets of data corresponding to smaller $\sqrt{\tau}$ cover 
a slightly smaller range in $x_F$, while at increasing 
$\sqrt{\tau}$ 
the event numbers filling each experimental bin decrease, 
and error bars increase. For comparison, in fig.2 we report 
data from table VI of \cite{Conway89} in a mass range 
near the upper edge 9 GeV/c: $\sqrt{\tau}$ ranges 
from 0.392 to 0.415 (mass between about 8.5 and 9 GeV/c$^2$). 
Clearly, the error bars are much bigger than in the case 
of fig.1. 
The corresponding simulated event numbers are extracted from 
a set of 100,000 sorted events between mass values 7 and 9.2 
GeV/c$^2$. 

Data from \cite{Conway89} do not cover $x_F$ $<$ $-0.1$ or 
$-0.2$. This is a typical situation in fixed target experiments, 
where $x_{projectile}$ $\sim$ 1 and $x_{target}$ $\sim$ 0. 

The data and simulations in figs.1 and 2 are integrated 
with respect to the transverse momentum of the dilepton 
pair. In figs. 3 and 4 I report distributions with 
respect to $P_t$, for two assigned ranges of $x_F$: 
0-0.1 (fig.3) and 0.6-0.7 (fig.4). Here data and 
simulated distributions are integrated with 
respect to the mass (equivalently, to $\tau$) over 
the mass range 4-9 GeV/c$^2$. The meaning of 
open and filled squares is the same as in figs.1 and 2. 
Data points come from the same experiment of \cite{Conway89}, 
(they are reported explicitly in \cite{HEPDATA}; in 
\cite{Conway89} figures on the $P_t-$distributions are present, 
but a table of values is not reported)

\subsection{Comparison with experimental data: NA10}

The two richest collections of events at the kinematics 
interesting here have been provided by the 
collaboration NA10, with negative pions of 194 and 286 
geV on Tungsten\cite{NA10_194,NA10_286}. For the beam 
at 194 GeV, the data may be found in the final table of 
\cite{NA10_194}, while the data relative to the upper energy 
beam have been taken from \cite{HEPDATA}, to which they 
have been sent as a private communication. 
This experiment did not publish transverse 
momentum distributions. In addition, the covered $x_F$ 
range tends to become rather narrow near the lower dilepton 
mass value 4 GeV/c$^2$, where most events concentrate. So 
the most interesting distributions are at larger mass values 
with respect to E615. 

\begin{figure}[ht]
\centering
\includegraphics[width=9cm]{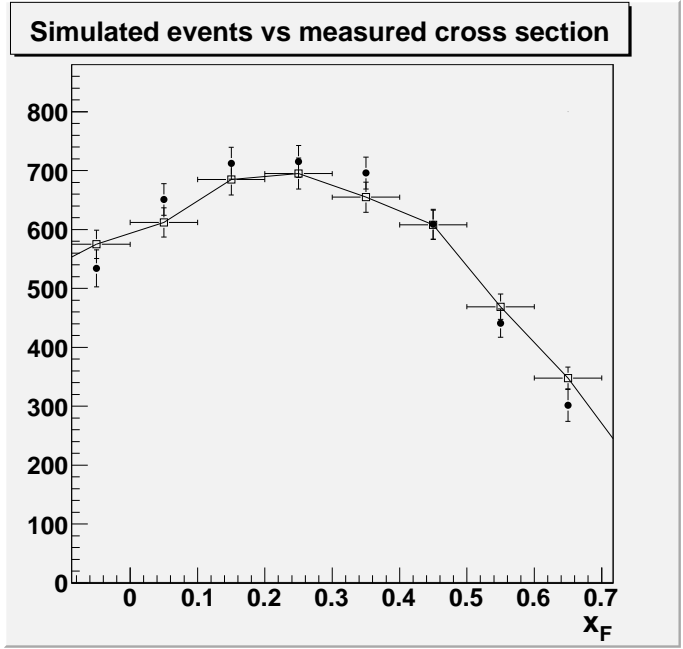}
\caption{
Filled squares: NA10 data at 194 GeV, for 
$\sqrt{\tau}$ in the range 0.33-0.36. 
Empty squares with joining line: Simulation (see text). 
\label{fig:X5}}
\end{figure} 

\begin{figure}[ht]
\centering
\includegraphics[width=9cm]{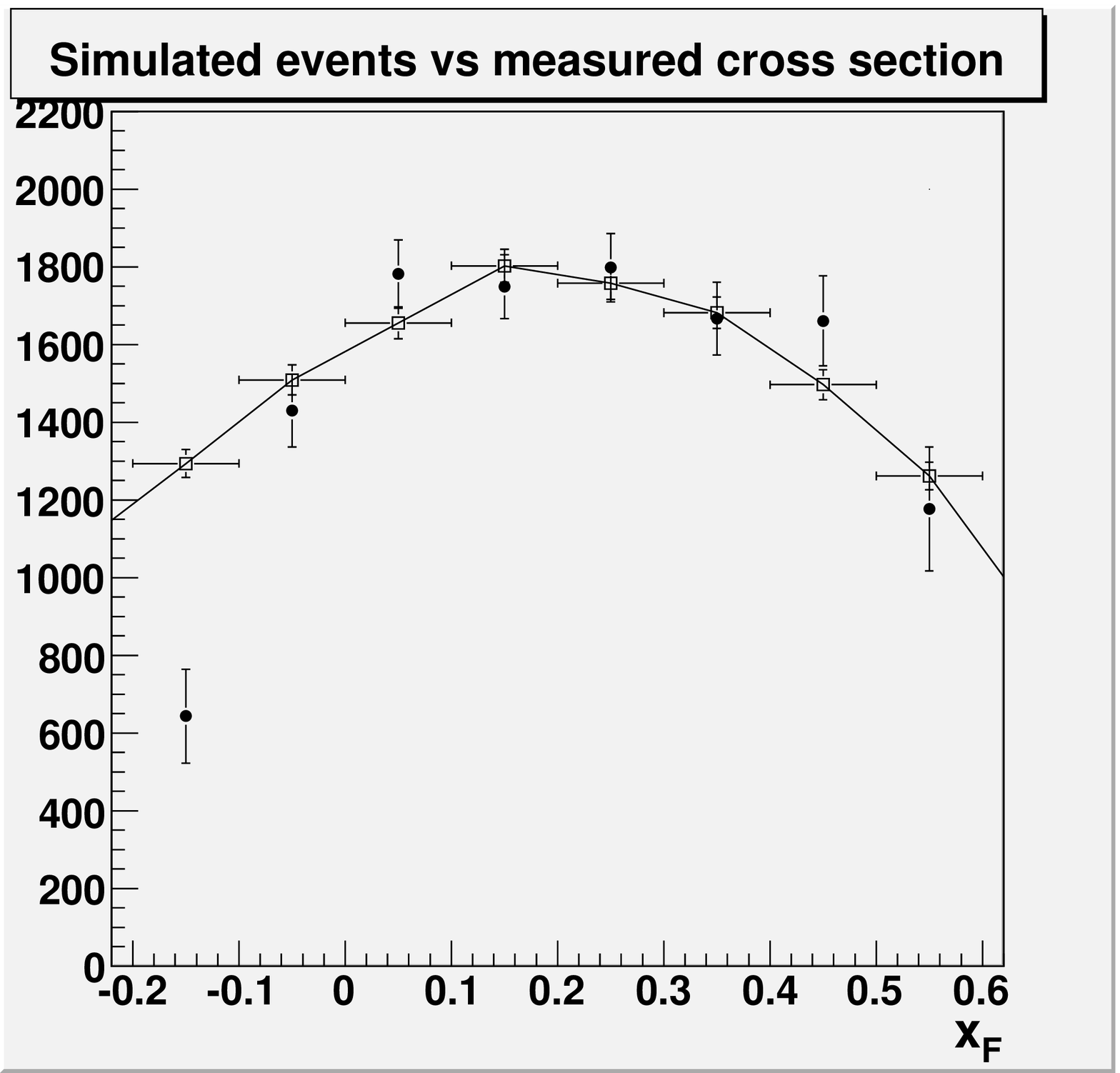}
\caption{
Filled squares: NA10 data at 286 GeV, for 
$\sqrt{\tau}$ in the range 0.33-0.36. 
Empty squares with joining line: Simulation (see text). 
\label{fig:X6}}
\end{figure}

\begin{figure}[ht]
\centering
\includegraphics[width=9cm]{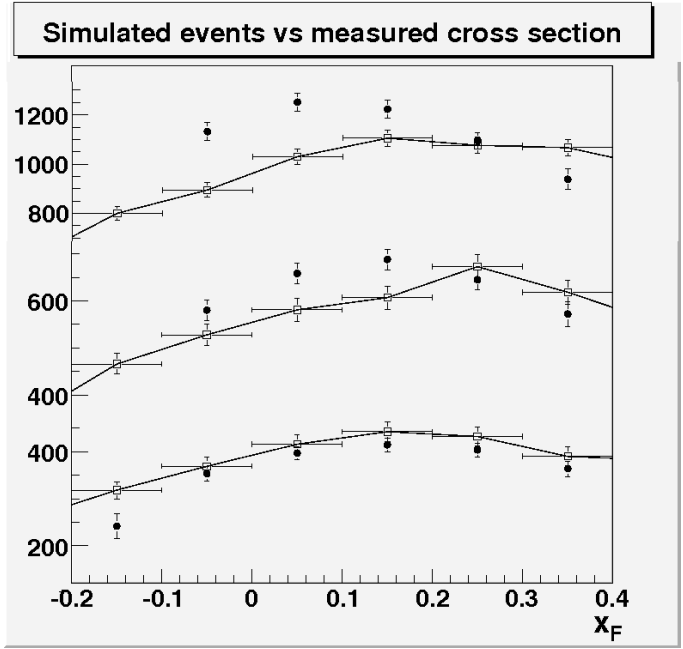}
\caption{
Filled squares: NA10 data at 286 GeV, for 
$\sqrt{\tau}$ in the ranges 0.21-0.24 (top), 0.24-0.27 (middle), 
0.27-0.3 (bottom). 
Empty squares with joining line: Simulation (see text). 
\label{fig:X7}}
\end{figure}

\begin{figure}[ht]
\centering
\includegraphics[width=9cm]{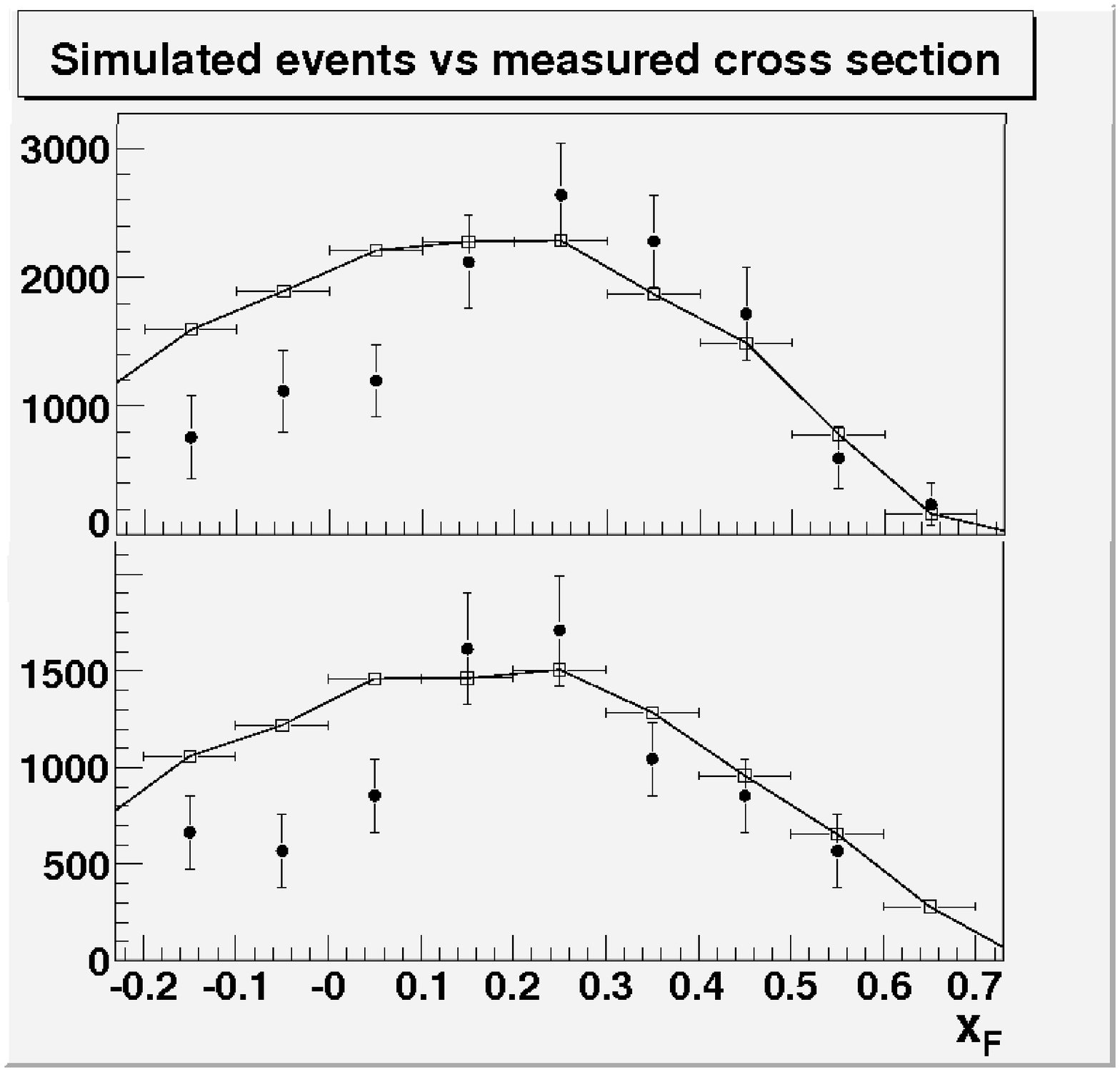}
\caption{
Filled squares: NA10 data at 286 GeV, for 
$\sqrt{\tau}$ in the ranges 0.51-0.54 (top) 
and 0.54-0.63 (bottom). These are equivalent to 
dilepton masses over the bottonium region. 
Empty squares with joining line: Simulation (see text). 
\label{fig:X8}}
\end{figure}

Data in fig.5 come from NA10 (194 GeV beam), and the 
simulated events are a subset of 100,000 events sorted by 
DY\_AB5 in the mass range 4-9 GeV/c$^2$, assuming a negative 
pion beam of 194 GeV hitting Tungsten. 
The data in figs. 6, 7 and 8 all refer to the upper energy beam 
of NA10, and their simulated counterparts have been 
extracted from a set of 100,000 events sorted between 4 and 7 
GeV/c$^2$, of 100,000 events between 7 and 9 GeV/c$^2$, 
and of 50,000 events between 11 and 15 GeV/c$^2$.

For the lower energy beam (pions of 194 GeV, meaning 
$s$ $\approx$ 364 GeV$^2$)
fig.5 reports the $x_F$ distribution 
for 
$\sqrt{\tau}$ in the range 
0.33-0.36, equivalent to dilepton mass between about 6.3 
and 6.9 GeV/c$^2$.

Fig.6 considers the same $\sqrt{\tau}$ range, for the case of the 
upper energy beam (286 GeV, meaning $s$ $\approx$ 
537 GeV$^2$). In this case 
this corresponds to dilepton mass between 7.6 and 8.3 GeV/c$^2$. 
According with the scaling hyppothesis, the two data distributions 
should be the very similar in the $x_F$ range 0-0.6 covered by 
both measurements.

For $\sqrt{\tau}$ $<$ 0.3, the $x_F$ range covered by NA10  
becomes small and data distributions are rather flat. 
Fig.7 reports data and simulations for the three $\sqrt{\tau}$ 
ranges 0.21-0.24, 0.24-0.27, 0.27-0.3, corresponding to 
masses ranging from 4.9 to 6.9 GeV/c$^2$. 

The compared examination of the three sets of fig.7 
suggests that (i) the simulation is worse for 
smaller $\sqrt{\tau}$, (ii) the $K-$factor extracted 
from E615 is slightly less steep (in its dependence on 
$\tau$) than the one extracted from NA10. 

The former fact 
depends on the increasing relevance of sea values 
for $x_{projectile}$ at small masses. The default distribution 
for $x_\pi$ in DY\_AB5 comes from E615 where sea 
(anti)quarks of the pion play a minor role in the fit of 
the pion distributions. For this reason, data from both 
NA10 and E615 
are reasonably fitted for $x_F$ $>$ $-0.1$ with the exception 
of small-$\sqrt{\tau}$ distributions. 

A signal of the 
relevance of sea partons from the pion side is the shift 
of the $x_F-$distribution peak towards $x_F$ $=$ 0. 
Valence-dominated measurements show this peak at 
$x_F$ $=$ 0.1-0.3. When the sea of the pion becomes relevant,  
it is probably better to substitute the default sea pion 
distribution of DY\_AB5 with more recent ones. Fig.7 suggests 
that 
this may be the case for $\sqrt{\tau}$ below 0.25. 

An interesting feature of NA10 is the presence of a conspicuous 
set of data at masses over the bottonium mass. This is $not$ 
the situation for which DY\_AB5 has been thought, however it 
is interesting to try and simulate these events. Fig.8 reports 
data and simulation for $\sqrt{\tau}$ in the ranges 0.51-0.54 
(mass between 11.8 to 12.5 GeV/c$^2$) 
and 0.54-0.63 (mass between 12.5 and 14.6 GeV/c$^2$). 

We see that DY\_AB5 has difficulties in reproducing  
the shape of these event 
distributions for $x_F$ $<$ 0.2. Actually, the ``gap'' 
of these data distributions at $x_F$ $\approx$ 0 looks 
a little unnatural. More in general, in all the previous 
figures the agreement between montecarlo and data is worse 
for negative $x_F$, where data distributions fall rather 
steeply. This could be related with the fact 
that negative $x_F$ data are at the border of the 
regions of good acceptance for fixed target experiments.

\subsection{Sivers asymmetry plots in different calculation schemes}

As observed in Section 2.5, 
the event cross section may be written in the form 
$\sigma$ $=$ $\sigma_0(X_1,X_2,P_t,\theta)$ 
$[1+A(X_1,X_2,P_t,\theta,..)]$, where the former term expresses 
that part of the cross section that does not contain azimuthal 
and spin asymmetries, while the asymmetries themselves are contained 
in $A$, and where $A$ may be approximated in different ways. 
In particular, in DY\_AB5 $A$ contains flavor weight factors, 
that are absent a more recent version DY\_AB6. In  
DY\_AB6 the full cross section $\sigma$ is a sum of independent terms each 
referring to a sea or valence parton. 
Each flavor contribution carries ``its own'' asymmetry. 
In DY\_AB5 
$\sigma_0$ is a sum of flavor contributions, and $A$ is an independent
sum of flavor contributions (see\cite{BR_JPG2}), weighted with 
good sense.

\begin{figure}[ht]
\centering
\includegraphics[width=9cm]{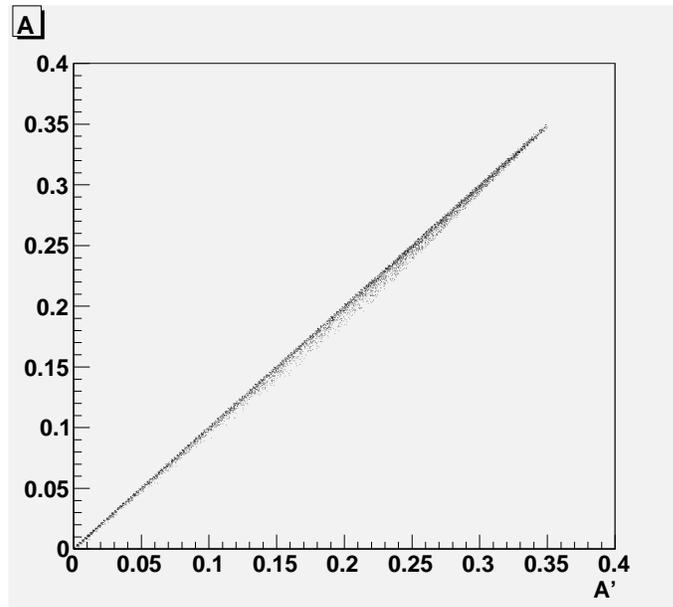}
\caption{
Scatter plot of asymmetries calculated by DY\_AB5 (horizontal) 
and by DY\_AB6 (vertical), for 7000 sorted $\pi^-p$ Drell-Yan 
events at $s$ $=$ 100 GeV$^2$, in the mass range 4-9 GeV/c$^2$. 
From \cite{BR_JPG2}. 
\label{fig:X9}}
\end{figure}

\begin{figure}[ht]
\centering
\includegraphics[width=9cm]{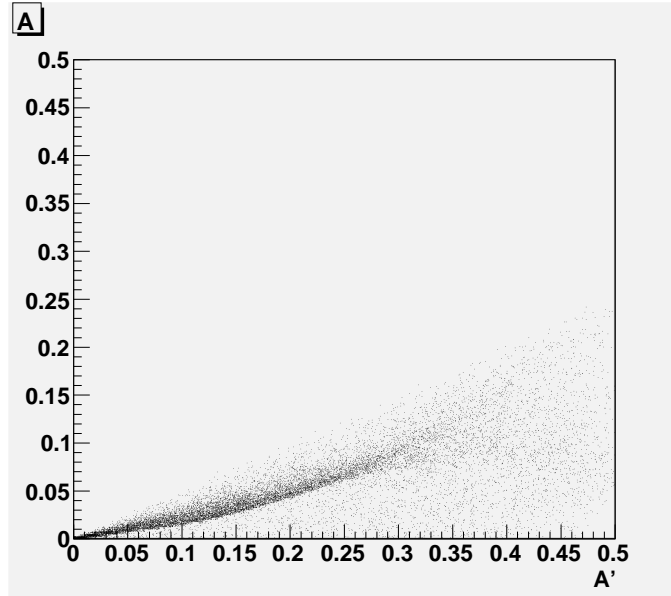}
\caption{
Scatter plot of asymmetries calculated by DY\_AB5 (horizontal) 
and by DY\_AB6 (vertical), for 20000 sorted $\pi^+p$ Drell-Yan 
events at $s$ $=$ 100 GeV$^2$, in the mass range 1.5-2.5 
GeV/c$^2$. From \cite{BR_JPG2}. 
\label{fig:X10}}
\end{figure}

As far as azimuthal/spin asymmetries are neglected, the 
two codes produce the same results ($\sigma_0$ is the same 
in both cases). When asymmetries are included, 
the scheme implemented by DY\_AB6 is more proper. 

Exploiting the equality of $\sigma_0$ in the two cases, in 
\cite{BR_JPG2} Drell-Yan events have been sorted according 
with $\sigma_0$ only, and the corresponding Sivers asymmetry 
(deriving from the $A$ term) has been calculated within 
the relations of DY\_AB5 and of DY\_AB6. The 
the scatter plot of the two asymmetry calculations for each 
event are reported in several figures. From that work two figures 
are borrowed, showing the two most ``extreme'' cases. 

Fig.9 reports 7000 events for $\pi^-p$ Drell-Yan at $s$ $=$ 
100 GeV$^2$, lepton invariant mass in the range 4-6 GeV/c$^2$, 
transverse momentum in the range 1-3 GeV/c. 
The Sivers asymmetry is parameterized according to \cite{BRMCc}.  

Fig.10 reports 20000 events for $\pi^+p$ Drell-Yan at $s$ $=$ 
100 GeV$^2$, lepton invariant mass in the range 1.5-2.5 GeV/c$^2$, 
transverse momentum in the range 1-3 GeV/c. 
The Sivers asymmetry is parameterized according to \cite{Torino05}.  

In the former case DY\_AB5 and DY\_AB6 would produce the same 
Sivers asymmetry. In the latter the difference is striking. 
As deeply discussed in \cite{BR_JPG2} the difference between the 
two is proportional to the role of sea (anti)quarks. Positive 
pions 
on protons, and small dilepton mass, enhance the role of 
sea partons. 

For these reasons, the code DY\_AB6 has been prepared, aiming 
at situations where sea partons could become more and more 
relevant. 

However, the use of DY\_AB6 has been up to now restrained by the 
fact that the available parameterizations of functions like the 
Sivers one are normally produced by fitting data within the same 
``ideological'' scheme of DY\_AB5. In these cases, the use of 
DY\_AB6 would increase errors, instead of decreasing them. 
This is discussed in \cite{BR_JPG2} with plenty of details. 

The main point is that when a ``global fit'' is undertaken 
using a scheme like the one of DY\_AB5, the effect of sea 
partons is effectively included inside valence quark 
distributions. So, when the results from such a global 
fit are emploied in DY\_AB6 for predicting some result, it is 
very dangerous to add separate sea quark contributions that are 
already present in valence quark distributions. 

The place where DY\_AB6 can be more proper is the one 
of $theoretical$ models for the Sivers function,  
built according with the 
scheme where each flavor is individually considered. 
But for the aim of modelling an experiment apparatus, 
one normally prefers using phenomenological parametrizations, 
rather than theoretical models.

\end{document}